\documentclass[aps,prx,reprint,showpacs,superscriptaddress,floatfix]{revtex4-2}
\usepackage[utf8]{inputenc}
\usepackage{graphicx}
\usepackage{ulem}
\usepackage{hyperref}
\hypersetup{
colorlinks=true,final=true,
        linkcolor=blue,
        citecolor=blue,
        filecolor=blue,
        urlcolor=blue,
}

\begin{document}

\title{Role of correlations in Ruddlesden-Popper bilayer nickelates under compressive strain}

\author{Logan Bleys}
\affiliation{Department of Physics, Kent State University, Kent, OH 44242, USA}

\author{Nicholas Corkill}
\affiliation{Department of Physics, Kent State University, Kent, OH 44242, USA}

\author{Yi-Feng Zhao}
\affiliation{Department of Physics, Arizona State University, Tempe, AZ 85287, USA}

\author{Gheorghe Lucian Pascut}
\affiliation{MANSiD Research Center and Faculty of Forestry, Stefan Cel Mare University (USV), Suceava 720229, Romania}

\author{Harrison LaBollita}
\affiliation{Center for Computational Quantum Physics, Flatiron Institute,
162 5th Avenue, New York, New York 10010, USA}

\author{Antia S. Botana}
\email{antia.botana@asu.edu}
\affiliation{Department of Physics, Arizona State University, Tempe, AZ 85287, USA}

\author{Khandker F. Quader}
\email{quader@kent.edu}
\affiliation{Department of Physics, Kent State University, Kent, OH 44242, USA}

\begin{abstract}

The recent discovery of superconductivity in thin films of the bilayer Ruddlesden-Popper (RP) nickelate La$_3$Ni$_2$O$_7$ (La327) under compressive strain has generated enormous interest,
opening up further opportunities to stabilize superconductivity in this class of materials at ambient pressure.
To better understand the many-body normal state from which superconductivity arises, it is important to ascertain the nature and role of correlations in its electronic structure.
 To provide insights into this question, we use a fully charge self-consistent DFT+e-DMFT (eDMFT) approach to study La327 at several compressive strain levels. 
 At the strain level where superconductivity has been observed experimentally (-2\%), in contrast with DFT and DFT+$U$ results, the so-called $\gamma$ pocket emerges and the associated band, of mostly $d_{z^2}$ character, crosses the Fermi level exhibiting `flat band''-like features when dynamical correlations are included. Larger strain levels
suppress the $\gamma$ pocket, which may have implications for superconductivity or its pairing symmetry.

\end{abstract}

\maketitle

\section{Introduction}
\label{sec_Introduction}

 Over the several decades following the discovery of high-temperature superconducting cuprates \cite{Bednorz1986}, there have been sustained efforts
 to find other classes of high-T$_c$ superconductors \cite{Norman_2016}. 
 Nickelates constitute the most recent class of materials that have contributed to exciting developments in this arena \cite{mitchell_review, frontiers_es}.
 Since the discovery of superconductivity in thin films of hole-doped infinite-layer  nickelates,  
 $R$NiO$_2$ ($R$ = rare earth Nd, Pr, La) in 2019 \cite{Li-Nature19,Osada-Pr-Nano20,Zeng-La-SciAdv}, 
rapid progression in materials synthesis resulted in the observation of superconductivity in the $n=5$ member of the same structural family  Nd$_6$Ni$_5$O$_{12}$ \cite{pan2021super}. In 2023, superconductivity was reported in the $n=2$ (La$_3$Ni$_2$O$_7$) and $n=3$ (La$_4$Ni$_3$O$_{10}$) members of the parent Ruddlesden-Popper (RP) series R$_{n+1}$Ni$_n$O$_{3n+1}$  under pressure ~\cite{sun2023superconductivity, wang2024bulk, hou2023emergence, zhang2023superconductivity, li2023signature, zhu2024superconductivity}, with the highest observed $T_c$ being $\sim$ 80 K in the bilayer compound. Recently, superconductivity under epitaxial compressive strain in the $n=2$ RP nickelates has also been reported ~\cite{Hwang25,Zhou25,Liu25} with a $T_c$ $\sim$30 K - 40 K,  for a strain level $\sim$ $-$2\% provided by a LaSrAlO$_4$ (LSAO) substrate. This latest finding should enable further opportunities to stabilize and study superconductivity in RP nickelates at ambient pressure.

\begin{figure*}
    \centering
    \includegraphics[width=1.0\linewidth]{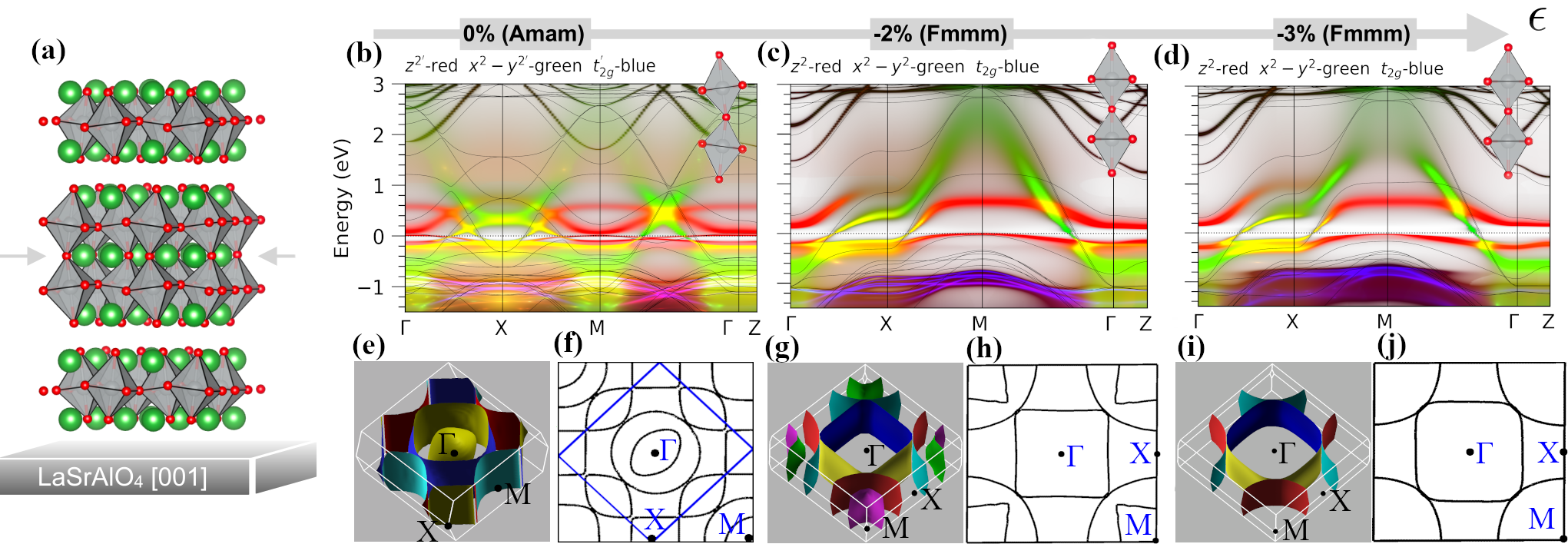}
        \caption{{\bf Evolution of the orbital resolved spectral function and Fermi surface for La327 under compressive strain ($\epsilon$).} Panel (a) shows the La327 crystal structure along with the LSAO substrate. Panels (b), (c), and (d) show our calculated eDMFT orbital resolved spectral function for an {\it Amam} structure at 0\% strain; for an {\it Fmmm} structure at -2\% and -3\% strains, respectively. The calculated DFT band structures are overlaid on the eDMFT spectral functions and shown as thin lines. Panels (e), (g), (i) show the 3D interacting Fermi surfaces obtained within eDMFT for 0\% ({\it Amam}), -2\%, and -3\% ({\it Fmmm}) strain levels,
       respectively. Panels (f), (h), (j) show the $k_z= 0$ 2D slices of the 3D Fermi surfaces in (e), (g), (i), respectively. The blue diamond in panel (f) denotes the reduced (or {\it Fmmm} zone); note the zone folding of {\it Amam} relative to {\it Fmmm}. Primed orbitals are used in the {\it Amam} structure as they are defined in the local coordinate system of the (rotated) octahedra. The results are for  $U$=7 eV, $J$=1 eV at 300 K.}
    \label{fig1}
\end{figure*}
 
 The discovery of superconductivity in strained La327 films, however, has also posed new fundamental questions about whether the nature and origin of superconductivity are the same under pressure and strain. In the context of the electronic structure, a pivotal question is if an extra Fermi surface (FS) sheet (the so-called $\gamma$ sheet), associated with the bonding Ni-$d_{z^2}$ band, exists in the strained bilayer \cite{luo2023bilayer, Zhang2023, Yang2023, Yang2024, Zhang2024save, sakakibara2024, christiansson2023correlated}. 
 This relates back to the emergence of such FS pocket in bilayer La327 under pressure and its possible connection to the observed superconductivity with dominant $s^\pm$ pairing \cite{lechermann2023electronic, Yang2024,Zhang2024,Zhang2024save, zhang2024structural, Lu2024}.
 A number of density functional theory (DFT) and DFT + $U$ ($U$= on-site Coulomb repulsion) works~\cite{Zhao25,Geisler25,Yi25,Regan25,rhodes2024structural} on the electronic structure of La327 under strain find that this $\gamma$ band is shifted down in energy below the Fermi level (in contrast to the results obtained under pressure); with the effect of a static $U$ actually being to further push it to lower energies.
 The experimental electronic structure, vis-à-vis angle-resolved photoemission spectroscopy (ARPES), is not yet settled, with one work finding the $\gamma$ band below ($\sim$ 70 meV) the Fermi energy and another showing that it actually crosses it, likely due to hole doping effects ~\cite{ARPES1,ARPES2}.  
 
 In order to have a better understanding of the electronic structure and emergent properties of La327 under strain, it is important to further scrutinize the role of electronic correlations in its many-body normal state. In particular, it is relevant to ascertain if the physics of this bilayer nickelate is governed by single- or multi-orbital degrees of freedom. To shed light on these issues, in this work we study bilayer La327 at several compressive strain levels using a fully charge self-consistent first principles DFT+e-DMFT method (hereafter referred to as eDMFT)~\cite{Georges-Kotliar-DMFTmethod-RMP1996,Kotliar-Haule-DMFTmethod-RMP2006,Haule-DMFTfullpot-PRB2010,Haule-DMFTdblecount-PRL15}.  We find that the electronic structure including dynamical correlations has important differences when compared to that obtained at the (static) DFT+$U$ level: the $\gamma$ Fermi surface sheet that is well below the Fermi level in DFT+$U$ under a 2\% compressive strain is recovered in eDMFT. Overall, both the $d_{x^2-y^2}$ and $d_{z^2}$  orbitals cross the Fermi level, and play active roles in the electronic structure of La327 at a 2\% compressive strain. Larger strain levels do suppress the $\gamma$ pocket, which may have implications for superconductivity and/or its pairing symmetry. 
 Our calculations show that, unlike the $d_{x^2-y^2}$ orbital, the $d_{z^2}$ states do not attain full coherence until a low temperature is reached.

\section{Results} 

\textbf{Crystal structure and computational approach.} 
Our results take into account recently reported structural changes in La327 under strain \cite{bhatt2025resolving}. We start by putting these changes into context with respect to structural trends under pressure wherein La327 has been shown to undergo a structural transition from
orthorhombic $Amam$ to $Fmmm$ space groups at $\sim$ 14 GPa, concomitant with the emergence of superconductivity \cite{sun2023superconductivity,hou2023emergence,wang2024bulk,zhu2024superconductivity}. This
transition suppresses the octahedral tilts 
as reflected in the change of the Ni-O-Ni bond angle
across the apical oxygens from 168$^{\circ}$
to 180$^\circ$. Compressive strain indeed mimics this effect of straightening the apical bond angle, turning the structure of La327 into $Fmmm$-like, while the octahedra become further buckled under tensile strain with the structure remaining $Amam$-like \cite{bhatt2025resolving}. The structural trends observed in thin films are corroborated by DFT calculations \cite{Zhao25,Geisler25}.  As a consequence, we study compressive strain levels of $-2$\% and $-3$\% in $Fmmm$ symmetry, and the unstrained case in $Amam$ symmetry. 
We have used as a starting point the structures generated by us in Ref. \cite{Zhao25}, whose details can be seen in Table 1 of Supplementary Note 1.  

Our fully charge self-consistent eDMFT calculation procedure applied to compressively strained and unstrained La327  is 
provided in the Methods section.
Localized orbitals for the Ni-$3d$ states are constructed from a large hybridization window $\pm 10$ eV around the Fermi level. A five-orbital Slater Hamiltonian governs the interactions in the quantum impurity problem. The Coulomb tensor is parameterized by a $U$ = 7 eV and a Hund's coupling $J$ = 1 eV, motivated by previous constrained DMFT results~\cite{Pascut23}. For the results in the main text, we choose a system temperature of $T = 300$ K. We provide additional results at a lower system temperature ($100$ K) in Supplementary Note 2. All calculations are performed enforcing paramagnetism.

\textbf{Evolution of the orbital resolved spectral function and Fermi surface topology with strain.}
Figure \ref{fig1} shows the evolution of the DFT and eDMFT electronic structure of La327 with strain. To match the experimental structural conditions~\cite{bhatt2025resolving}, we study La327 in \textit{Amam} symmetry at zero strain and in \textit{Fmmm} symmetry for the $-2\%$ and $-3\%$ strain states, as mentioned above.  
The La327 crystal structure is shown in Fig. \ref{fig1}a. We start by describing the electronic structure for 0\% strain. The DFT band structure using the generalized-gradient approximation (GGA) as the exchange-correlation functional has been shown in previous work \cite{Yang2023, lechermann2023electronic} and is characterized near the Fermi level ($E_F$) by the presence of Ni-$e_g$ ($d_{z^2}$ and
$d_{x^2-y^2}$) states, 
indicating that these two Ni-$d$ orbitals play the dominant
role in the low-energy electronic structure of this material. This is expected given the nominal Ni valence in the
bilayer RP of 2.5+ (corresponding to $d$$^{7.5}$). This average
$d$ filling implies that 1.5 e$_g$-electrons per Ni (3 per bilayer)
need to be accommodated close to the Fermi level, while
the $t_{2g}$ orbitals are filled. The Ni-$d_{z^2}$ states are split
by $\sim$ 1 eV into a bonding and antibonding molecular orbital combination due to the quantum confinement of
the nickel-oxygen bilayers in the structure~\cite{pardo2010,pardo2011}. The bonding $d_{z^2}$ band crosses the Fermi level in unstrained La327 at the GGA level. Instead, static (Hartree) correlations at the DFT+$U$ level push this band to lower energies \cite{Yang2024arpes}. This is unsurprising given that the DFT+$U$ mechanism is designed to occupy (deoccupy) partially filled bands.  Comparing the DFT band structure (thin line overlay in Fig. \ref{fig1}b) and 
the eDMFT orbital-resolved spectral function, dynamical correlations do not simply produce a rigid band shift like DFT+$U$. There are clear band renormalization effects: both the antibonding and bonding $d_{z^{2}}$, as well as the  $d_{x^2-y^2}$  bands are pushed towards the  Fermi level from above and below, notably at the high-symmetry points X and M, and along M-$\Gamma$ high-symmetry path. The color scheme in Fig. \ref{fig1}b is a measure of the degree of $e_g$ orbital hybridization, which can be seen to be large near the Fermi energy.
 Importantly, our eDMFT calculations do not suppress the $\gamma$ pocket, unlike DFT+$U$, and this can also be seen in the corresponding Fermi surface (Fig. \ref{fig1}e, f). While ARPES experiments \cite{ARPES1} do not show this band crossing the Fermi level, it is very close to it, with its exact position likely being very sensitive to stoichiometry issues.  

Under a 2\% compressive strain (the strain level at which superconductivity has been experimentally observed), the electronic structure at the DFT(GGA) level shows a clear shift of the bonding $d_{z^2}$ band below the Fermi level with respect to the unstrained case. The inclusion of dynamical correlations in the Ni-$d$ electrons gives rise to important changes in the low-energy physics. As the eDMFT orbital resolved spectral function shows (Fig. \ref{fig1}c), relative to the DFT case (thin lines in the same panel), the bonding-antibonding splitting gets drastically reduced. Both Ni-$d_{z^2}$ bands move towards the Fermi level, and the bonding $d_{z^2}$ crosses it. This is in clear contrast with the DFT+$U$ results~\cite{Geisler25,Regan25} that show that this band is shifted even further away from the Fermi level with respect to the GGA case. A sizable decrease of the bandwidth of the $d_{z^2}$ bands can also be observed. For example, the bandwidth of the upper (antibonding) band can be seen to decrease from $\sim$ 2 eV to 0.75 eV.  The lower (bonding) $d_{z^2}$ band shows a pronounced ``flat-band"  feature at $E_F$ along the high-symmetry $k$-path X-M-$\Gamma$. 
Figure~\ref{fig1}c also depicts the large degree of $e_g$ hybridization around the Fermi level.  Consistent with our eDMFT orbital resolved spectral function, the eDMFT Fermi surface  (Fig. \ref{fig1}g, h) shows an extra corner sheet ($\gamma$ sheet)
compared to the DFT Fermi surface. Thus, at the eDMFT level, both the $d_{z^2}$ and $d_{x^2 - y^2}$ bands cross $E_F$ in La327 under a 2\% compressive strain, and play an active role in its low-energy physics. Based on these results, multi-orbital Ni-$e_{g}$ physics is likely relevant for the observed superconductivity and the corresponding pairing symmetry.  Such a multi-orbital scenario appears to be different from the cuprates. 
We finally focus on the effect of increasing the degree of compressive strain on the electronic structure.  Fig. \ref{fig1}i, j
shows how the size of the $\gamma$ pocket progressively decreases with increasing strain, disappearing at a -3\% strain level. 
Consistent with this Fermi surface, the orbital resolved spectral function for a -3\% strain (Fig. \ref{fig1}d) shows how the $d_{z^2}$ bonding band is shifted down in energy to just below the Fermi level. 
We elaborate on the implications of these electronic structure changes in the Discussion section below. \\

\begin{figure}
 \centering
	\includegraphics[clip, width=0.9\columnwidth]{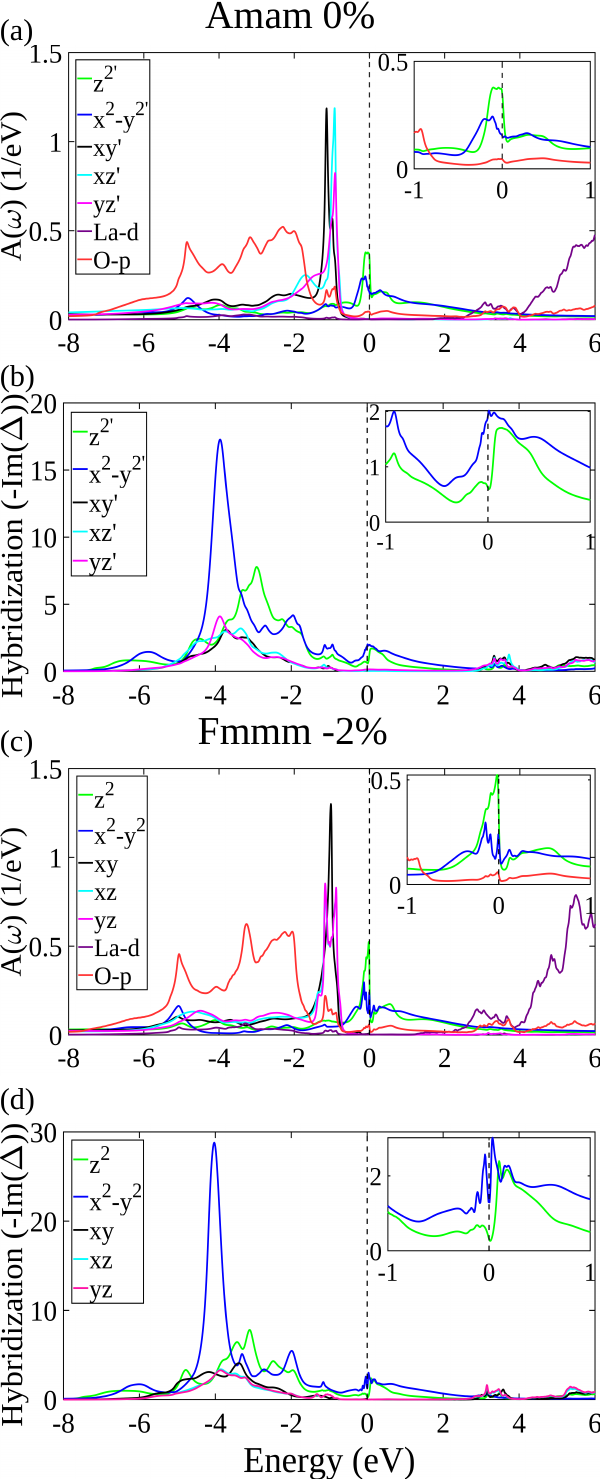}
	\caption{{\bf Evolution of the {\bf k}-integrated spectral function and hybridization for La327 under compressive strain}. Panels (a),(c)  show the spectral function (projected density of states, pDOS) per atom for the Ni-$d$, O-$p$ and La-$d$ orbitals for the {\it Amam} phase at 0\% strain, and for the {\it Fmmm} phase at -2\% strain, respectively. Panels (b),(d) show the hybridization of the Ni-$d$ orbitals with the bath electrons for the cases corresponding to (a), (c).  
    The insets focus on energy around the Fermi level, which is marked by zero on the energy axis. Primed orbitals are used in the $Amam$ structure as they are defined in the local coordinate system of the (rotated) octahedra.}
	\label{dos}
\end{figure}

\textbf{eDMFT density of states and hybridization functions.} Fig. \ref{dos} shows the behavior of 
the ${\bf k}$-integrated eDMFT spectral function (partial density of states, pDOS) 
and the hybridization of the correlated Ni-$d$ orbitals with the bath electrons for a 0\% strain level ($Amam$ symmetry) and a  2\% compressive strain  ($Fmmm$ symmetry). The results at a 3\% compressive strain are similar to those of a 2\% and are shown in Supplementary Fig. 2a,b of Supplementary Note 3. In all cases, the pDOS per atom is shown for all the Ni-$d$, O-$p$, and La-$d$ orbitals at 300 K (Fig. 2a, c).
For all strain levels, 
the pDOS center of mass for
the Ni-$d_{x^2-y^2}$ and Ni-$d_{z^2}$ orbitals is around the Fermi level, while for the $t_{2g}$ orbitals, the centroid is at lower energies. 
The bonding $d_{z^2}$ flat band can be observed at the Fermi level in both cases. For all three strain levels, there is a clear admixture between $d_{x^2-y^2}$, $d_{z^2}$, and O-$2p$ states at the Fermi level.  While the Ni-$d$ states hybridize strongly with the O-$p$ states, their overlap with  
La-$d$ states is negligible. Fig. \ref{dos}b,d further shows that at the Fermi level, the hybridization of Ni-$d_{x^2-y^2}$ states is much stronger than that of Ni-$d_{z^2}$, so that the Ni-$d_{z^2}$ orbital maintains more of its pure character. This is also reflected in the orbital projected spectral function plots described in the previous section. 
The pDOS for the $d_{z^2}$ orbital at $-$2\% strain and $-$3\% strain (Supplementary Fig. 2a of Supplementary Note 3) have a larger width around the Fermi level, compared to that of $d_{x^2-y^2}$. 
Taken together with the orbital resolved spectral function at $E_F$, and the behavior of the self-energy (discussed below), 
this is indicative of the $d_{z^2}$ orbital not attaining  full coherence  at 300 K, while $d_{x^2-y^2}$ displays coherent quasiparticle behavior. 
In the 0\% \textit{Amam} case, both the $d_{z^2}$ and $d_{x^2-y^2}$ orbitals display larger widths in the pDOS around $E_F$ and hence none of them attains full coherence at this temperature. %Again, this can be more clearly seen in 
We note that our DMFT calculations give an occupation for the Ni-3$d$ electrons, $n_d(U)$ ($U$=7 eV) $\sim$8.2, which is larger than the nominal value of 7.5. These occupation numbers are similar to those obtained in prior DMFT work in La327 under pressure \cite{lechermann2023electronic}.

\begin{figure}
 \centering
	\includegraphics[clip,width=0.9\linewidth]{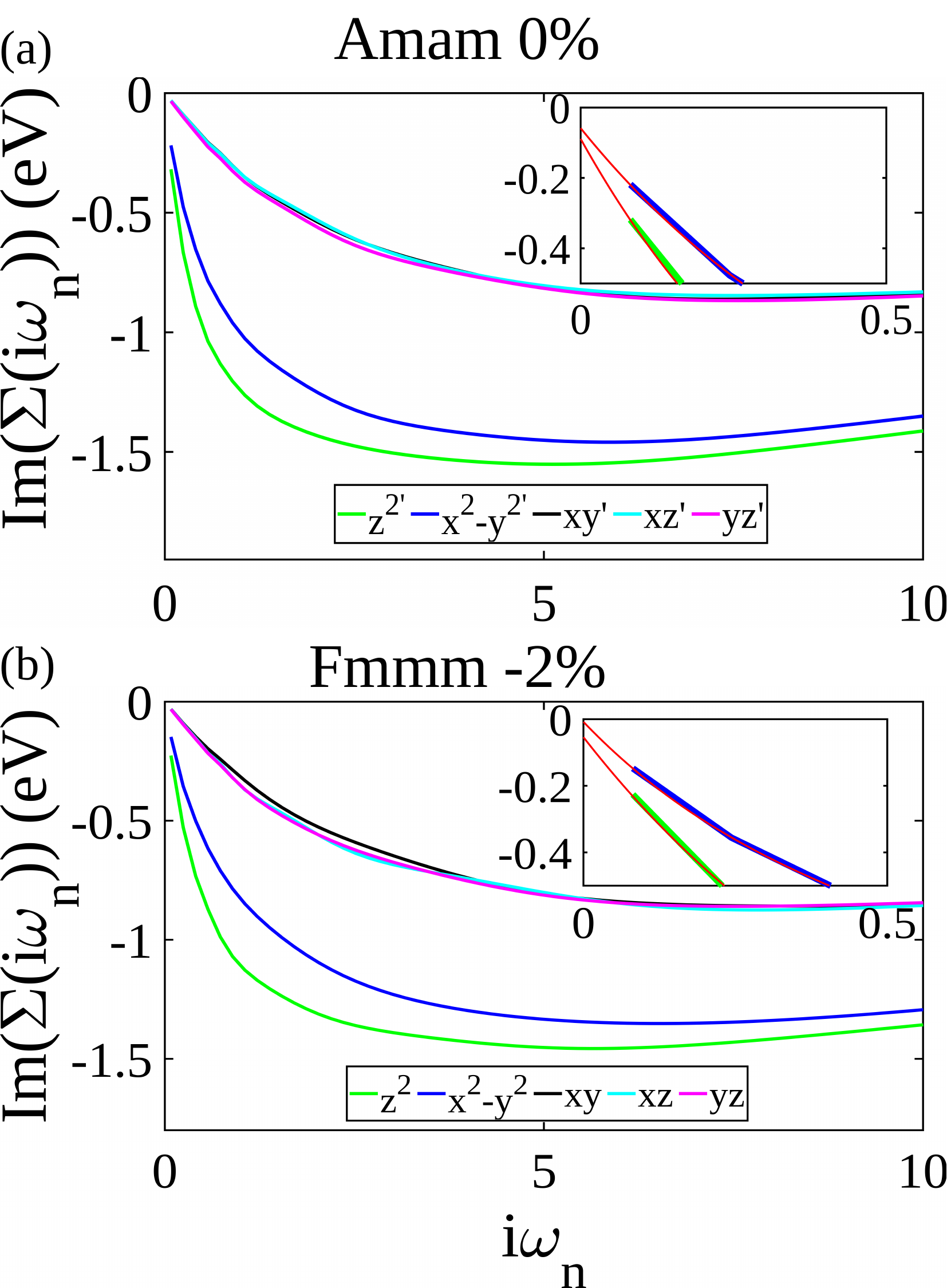}
    \caption{\textbf{Evolution of the imaginary self-energy of the Ni-$d$-electrons on the Matsubara frequency axis for La327 under compressive strain.} Panel (a) shows the result for the unstrained $Amam$ structure; panel (b) for the $Fmmm$ structure at -2\% strain. Primed orbitals are used in the $Amam$ structure as they are defined in the local coordinate system of the rotated octahedra. To extrapolate to zero Matsubara frequency, we performed polynomial fits to our self-energy curves (red curves).  }
    \label{selfenergy}
\end{figure}

\textbf{Self-energies.} Figure~\ref{selfenergy}a,b shows the imaginary part of the self-energy of the Ni-$d$ electrons on the Matsubara frequency axis for 0\% strain ($Amam$) and for $-$2\% strain ($Fmmm$) at 300 K. To explore the effect of temperature on scattering, in the $-2$\% strain case, we performed calculations also at 100 K (the latter are shown in 
Supplementary Fig. 1 of Supplementary  Note 2).  
For  $-$2\% strain, the 
imaginary self-energy for the $d_{z^2}$ orbital does not go to zero in the limit of zero Matsubara frequency, thereby pointing towards a sizeable scattering rate at 300 K. On the other hand, the self-energy does go to zero for $d_{x^2-y^2}$ in the $-$2\% $Fmmm$ case at this temperature. 
At the lower temperature of 100 K, the self-energy for both $d_{z^2}$ and $d_{x^2-y^2}$ goes to zero at zero Matsubara frequency (Supplementary Fig. 1e in Supplementary Note 2)  indicating that, compared to $d_{x^2-y^2}$,  $d_{z^2}$ achieves coherence at a lower temperature. For 0\% strain ($Amam$ phase), the self-energy of neither orbital appears to approach zero in the limit of zero Matsubara frequency, indicating that none of the $e_g$ orbitals achieves coherence down to 300 K in this case. 

\section{Discussion}
\label{sec_Discussions}

We have shown that the treatment of correlations in a 
non-perturbative manner in strained La327 (with both correlations and effects of the medium (bath) treated dynamically) gives rise to results that are rather distinct from those obtained from DFT+$U$ calculations (wherein correlations are static in nature).   
In La327, a lot of attention has been paid to the emergence of the $\gamma$ pocket in the electronic structure, as in several works it is linked to a dominant $s^{\pm}$ pairing symmetry of the superconducting order parameter. While a recent study~\cite{Sakakibara25} finds $s^{\pm}$ pairing regardless of the presence of the $\gamma$ pocket, other work has pointed out that the suppression of the $\gamma$ pocket might stabilize $d$-wave pairing 
instead \cite{Yang2024,qimiao2025}. In DFT + $U$ studies at a $-$2\% strain level (where superconductivity has been experimentally observed), the $d_{z^2}$ bonding band is shifted down in energy away from the Fermi level. In our eDMFT study, we find this band to cross the Fermi energy, resulting in the emergence of a $\gamma$ sheet in the Fermi surface. In this manner, at a $-$2\% strain, both the $d_{z^2}$ and $d_{x^2-y^2}$ bands play a substantive role in the low-energy physics of La327 after dynamical correlations are taken into account, and thus, from our calculations, multi-band physics is expected. The placement of the bonding $d_{z^2}$ band is, however, highly sensitive to the strain level, with increasing strain  
shifting it down to lower energies, resulting in the disappearance of the $\gamma$ pocket at a $-3$\% strain. 
The consequential progression of the $d_{z^2}$ bonding band (and hence the $\gamma$ pocket) starting from 0\% strain to higher strain bears some semblance to a Lifshitz-type transition~\cite{Lifshitz60} in the iron pnictides, where pressure-driven topological changes in the bands and Fermi surface have been shown to result in the collapse of the tetragonal phase~\cite{Quader-Widom14}. 

No consensus has been achieved on the experimental electronic structure in compressively strained La327 films as determined by ARPES when it comes to the position of the bonding $d_{z^2}$ band and the associated $\gamma$ pocket. One experiment~\cite{ARPES1} finds a $\gamma$ pocket submerged just below the Fermi level ($\sim$ 70 meV),
 while another~\cite{ARPES2} finds it above the Fermi energy. It is worth noting that the first experiment \cite{ARPES1} is
on La$_2$PrNi$_2$O$_7$, so the Fermi level would be slightly higher than that of  La$_3$Ni$_2$O$_7$, the compound we consider here. The second work \cite{ARPES2} states that the system is hole-doped by Sr from the substrate, thereby lowering the Fermi level. Overall, our results are not inconsistent 
with ARPES and with the overall sensitivity of the Fermi level placement with respect to structural/stoichiometric details. 

While an analysis of superconducting pairing symmetries is beyond the scope of our work, the
evolution of the $\gamma$ pocket with strain that we find suggests that in the scenario in which superconductivity is connected with the presence of the $\gamma$ sheet, its occurrence may be restricted to a 
narrow range of strain. This may not preclude other scenarios, such as the occurrence of superconductivity at higher compressive strains, but possibly with a different pairing symmetry and mechanism. 

\section{Summary and Conclusions}
\label{sec_Conclusion}

We have employed a DFT+e-DMFT (eDMFT) computational
framework to study the evolution of the electronic structure of La327 under strain, in light of the recent discovery of superconductivity in 2\% compressively strained thin films. The electronic structure obtained upon considering dynamical correlations is different from that obtained when only static correlations are considered at the DFT+$U$ level. In particular, we find that upon a 2\% compressive strain (in an $Fmmm$-like structure), a flat band of pure $d_{z^2}$ character crosses the Fermi level, in analogy to the pressurized case. The flat $d_{z^2}$ band at the Fermi level co-exists with a $d_{x^2-y^2}$ band that is much more strongly hybridized. Calculated scattering rates, orbital resolved and momentum integrated spectral functions indicate that in the {\it Fmmm} phase, 
the $d_{z^2}$ orbital does not attain full coherence, while $d_{x^2-y^2}$ does so at 300 K. Lower temperature results indicate that the two 
$e_g$ orbitals may have different coherence temperatures.
With increasing compressive strain, the bonding $d_{z^{2}}$ flat-band is suppressed from the Fermi surface. 
Our results suggest that if superconductivity is related to the existence of the $\gamma$ sheet, it may occur only in a narrow window of compressive strain.
Experiments at different strain levels would help understand trends in superconductivity and features of the normal state in bilayer RP nickelates.

\section{Methods}
\label{Sec_Methods}

{\bf DFT}: To obtain the electronic structure of the strained bilayer nickelate La327 in both the {\it Fmmm}~\cite{sun2023superconductivity}  and {\it Amam} cases, we constrained the in-plane lattice constants
to the appropriate {\it biaxial} strain level, $\epsilon = \frac{a^*(b^*)}{a(b)}$  and relaxed both the out-of-plane lattice constant and the internal coordinates. 
We employed the Vienna ab initio Simulation Package (VASP) for the structural optimizations~\cite{Kresse93,Blochl94,Kresse96}.
The generalized gradient approximation Perdew-Burke-Ernzerhof (GGA-PBE) functional was used for the exchange and correlation potential~\cite{PBE}.
The energy cutoff was set to 500 eV, and a k-point mesh $8\times8\times8$ was used in the relaxation procedure for all the strains and phases. All the
ionic positions were fully optimized until the Hellman-Feynman force was lower than 1 meV/A.. 
Using the VASP-optimized strained structures, we then performed electronic structure calculations using the full-potential linearized augmented plane wave (FP-LAPW) method as implemented in the WIEN2k code~\cite{WIEN2K-Blaha2020}; we employed the GGA-PBE functional for the WIEN2k calculations as well.

The radii of the muffin-tin (MT) spheres were taken to be  2.39, 1.85, and 1.64 a.u. for La, Ni, and O, respectively, in the $Fmmm$ structure. For the {\it Amam} case, we used slightly different MT spheres, namely, 2.0, 1.8, 1.6 a.u. respectively for  La, Ni, O. The RMT x K$_{\rm max}$ was set to 7, where K$_{\rm max}$ is the cutoff value of the modulus of the reciprocal lattice vectors and RMT is the smallest MT radius.
To obtain the self-consistent charge density in each step, we used the
tetrahedron integration method with 6000 k-points in the full Brillouin zone (the densities of states were computed using
30000 k-points in the full Brillouin zone). During the self-consistency cycles, the energy and charge criteria were set to
0.00001 Ry and 0.00001 $e^-$. 

{\bf eDMFT}: We employed the self-consistent DFT + embedded DMFT (eDMFT) method~\cite{Georges-Kotliar-DMFTmethod-RMP1996, Kotliar-Haule-DMFTmethod-RMP2006,Haule-DMFTfullpot-PRB2010,Haule-DMFTdblecount-PRL15} developed at Rutgers. Our calculations are fully charge self-consistent between the DFT and DMFT parts, 
i.e., the charge is recalculated after a DMFT run and then iterated through the DFT and DMFT loops until charge convergence is achieved. We employed an ``exact-d double-counting"~\cite{Haule-DMFTdblecount-PRL15}, and used the  ``Full Coulomb" option (as opposed to 'Ising'), i.e., the spin-rotation invariant form of the Coulomb repulsion of the Slater 
form.~\cite{Haule-DMFTfullpot-PRB2010} We used a large hybridization window of $\pm$ 10 eV. For the DMFT projectors, we chose quasi-atomic localized orbitals. The radial part is the solution of the Schroedinger equation in the muffin-tin sphere, with linearized energy at the Fermi level, and the angular dependence given by the spherical harmonics. In our projection technique, we only truncate the basis for the two-particle quantities, i.e., interaction and dynamic correlations are constrained to projection in real space around a given atom, while the large kinetic-part of the Hamiltonian is not approximated (or truncated) at all. All the degrees of freedom in this largest part of the Hamiltonian are treated by the complete basis. For the quantum impurity problem (Ni-$d$ electrons), we used a version of the continuous time quantum Monte Carlo (CTQMC) impurity solver. 
For the  Monte Carlo (MC) runs we employed at least 15 x 10$^6$ MC steps; the number of MC steps was
increased when lowering the temperature. We allowed the position of the chemical potential to vary during the self-consistent calculations.
The analytic continuation from the imaginary to real frequency axis was done using the maximum entropy method.
We note that for the $Amam$ structure, in view of the tilted octahedra (Fig. \ref{fig1}a), 
we rotated to the local coordinates to minimize the off-diagonal elements of the hybridization function to 
reduce the Monte Carlo sign problem. \\

{\bf Data availability}. The data that support the findings of this study are available
from the corresponding author on reasonable request.\\

{\bf Code Availability}: The DFT calculations were performed using the proprietary code VASP, WIEN2k, and 
open-source eDMFT code/impurity solver of Kristjan Haule at Rutgers University (http://hauleweb.rutgers.edu/tutorials/). 
Haule's eDMFT code is freely distributed for academic use under the Massachusetts Institute of Technology (MIT) License.

\section{Acknowledgements}
\label{sec_Acknowledgements}

A.S.B. and Y.Z. were supported by NSF Grant No. DMR-2323971 and acknowledge the ASU Research Computing Center for HPC resources.  G.L.P.’s work was supported by a grant of the Romanian Ministry of Education and Research, CNCS - UEFISCDI, Project No. PN-III-P1-1.1-TE-2019-1767, within PNCDI III. G.L.P. also acknowledges resources for the calculations provided by the STFC Scientific Computing Department’s SCARF cluster; preparation of the input files and data processing for the calculations were performed on the local cluster at Stefan cel Mare University of Suceava (USV), obtained through a grant of the Romanian Ministry of Education and Research, CNCS—UEFISCDI, project number PN-III-P1-1.1-TE-2019-1767, within PNCDI III.  K.F.Q.  acknowledges a QuantEmX grant from ICAM and the Gordon and Betty Moore Foundation, Grant GBMF5305, which partly funded this work. K.F.Q. also acknowledges computational resources for the calculations provided by Advanced Research Computing at Kent State University and the  Ohio Supercomputing Center. 
We thank M. Widom and M. J. Guttman for useful discussions. The Flatiron Institute is a division of the Simons Foundation.

\section{Author Contributions}
\label{sec_Auth-contrib}

A.S.B. and K.F.Q. conceived the project. N.C., L.B. G.L.P. and Y-F Z. performed the calculations. All authors processed and analyzed the data. K.F.Q. wrote the first draft. All authors discussed the results and arrived at the final version of the manuscript.\\

\section{Additional Information}
\label{sec_add-info}

{\bf Supplementary Information} accompanies this paper.\\

{\bf Competing interests:} The authors declare no financial or non-financial competing interests.\\

\newpage

\bibliographystyle{naturemag}

\bibliography{Bib_Nickelates.bib}

\vspace{0.3in}

\end{document}

\title{Supplementary Information:\\
Role of correlations in Ruddlesden-Popper bilayer nickelates under compressive strain}

\author{Logan Bleys}
\affiliation{Department of Physics, Kent State University, Kent, OH 44242, USA}

\author{Nicholas Corkill}
\affiliation{Department of Physics, Kent State University, Kent, OH 44242, USA}

\author{Yi-Feng Zhang}
\affiliation{Department of Physics, Arizona State University, Tempe, AZ 85287, USA}

\author{Gheorghe Lucian Pascut}
\affiliation{MANSiD Research Center and Faculty of Forestry, Stefan Cel Mare University (USV), Suceava 720229, Romania}

\author{Harrison LaBollita}
\affiliation{Center for Computational Quantum Physics, Flatiron Institute, 162 5th Avenue, New York, New York 10010, USA}

\author{Antia S. Botana$^*$}
\affiliation{Department of Physics, Arizona State University, Tempe, AZ 85287, USA}

\author{Khandker F. Quader$^*$}
\affiliation{Department of Physics, Kent State University, Kent, OH 44242, USA}

\begin{abstract}

\vspace{4.5in}

\rm{$^*$Corresponding Authors: Khandker F. Quader;\;Antia S. Botana}\\
\rm{Corresponding Author emails: quader@kent.edu;\;Antia.Botana@asu.edu}

\end{abstract}

\maketitle

\thispagestyle{empty}

\clearpage
\pagestyle{plain}

\begin{center}

\center{\bf Supplementary Note 1: STRUCTURAL PARAMETERS}\\
\vspace{0.3in}

Table \ref{tab:DFT spacing} shows the lattice constants, volume and relevant bond angles used in our calculations in the main text. 

\begin{table}[h!]
  \begin{center}
    \caption{DFT calculated lattice spacing and bond angle values.}
    \label{tab:DFT spacing}
    \begin{tabular}{c|c|c|c|c } % <-- Alignments: l for left, c for center, and r for right, with vertical lines in between
      \hline
      Strain ($\varepsilon$) & -3.0\% & -2.0\% & 0.0  \\  
     \hline
      Space group & $Fmmm$ & $Fmmm$ & $Amam$  \\
      Ni-planar O bond angle ($^{\circ}$)  & 2.0 & 0.9 & 9.4  \\
      Ni-apical O bond angle ($^{\circ}$)  & 0.0 & 0.0 & 11.4 \\
      In-plane axis (\AA) & 5.285/5.231 & 5.340/5.285 & 5.439/5.377  \\
      $c$ (\AA) & 20.92 & 20.73 &  20.40 \\
      Volume $a \times a \times c$\,(\AA$^3$) & 578.25 & 585.04 & 596.70 \\
      
      \hline
    \end{tabular}
  \end{center}
\end{table}

\end{center}

\setcounter{page}{1}

\vspace{0.3in}

\begin{center}

\center{\bf Supplementary Note 2: eDMFT RESULTS FOR -2\% STRAIN IN SYSTEM WITH {\it Fmmm} SYMMETRY AT 100 K} \\

\end{center}

\vspace{0.3in}

\parindent=0.2in

Here we present our e-DMFT results for {\bf k}-integrated spectral function, hybridization of the correlated Ni-$d$ electrons with bath electrons, orbital resolved spectral function and 3D Fermi surface  for -2\% in system with $Fmmm$ symmetry. These results show that there is no appreciable change in these quantities in going down from 300 K (in main text) to 100 K.
It may be noted that though there are no qualitative change in the location of  orbitals (spectral function), the spectral functions for $d_{x^2-y^2}$ and the bonding 
$d_{z^2}$ get sharper at the lower temperature close to the Fermi level. This reflects that these are becoming more coherent quasiparticles at this  temperature.  

\begin{figure*}[!htb]
\includegraphics[width=0.98\linewidth]{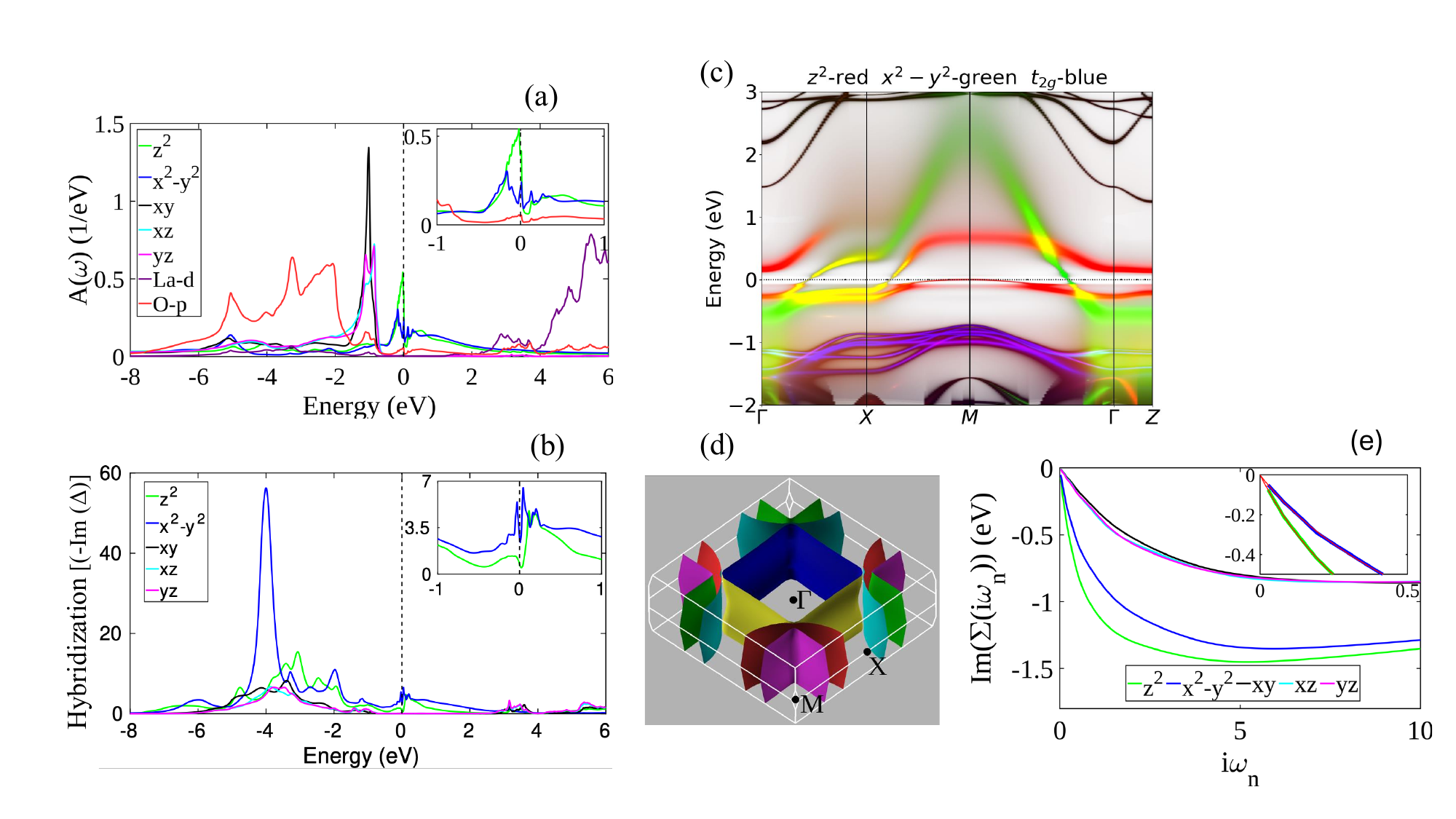}
\caption{\textbf{Results for -2\% strain {\it Fmmm} symmetry at 100 K}. Panels (a), (b) respectively shows the {\bf k}-integrated spectral function (partial density of states)) per atom
for all the Ni-$d$, O-$p$ and La-$d$ orbitals and hybridization of all Ni-$d$ orbitals with bath electrons. The insets are zoomed views of the two $e_g$ orbitals. 
Panel (c) shows the orbital resolved spectral function and panel (d) the 3D Fermi surface with the extra $\gamma$ sheet at the M-point, respectively. Panel (e) shows the Imaginary self-energy of Ni-$d$ orbitals on the Matsubara frequency axis; the inset shows that both the $e_g$ orbitals approach zero as zero frequency is approached at this temperature.
To extrapolate to zero Matsubara frequency, we performed polynomial fits to our self-energy curves (red curves).  Calculations were done with U=7 eV, J= 1 eV.}

\end{figure*}

\newpage

\begin{center}

\center{\bf Supplementary Note 3: e-DMFT RESULTS FOR -3\% STRAIN WITH $Fmmm$ SYMMETRY}\\

\end{center}

\vspace{0.3in}

\parindent=0.2in

Here we present our e-DMFT results for partial density of states ({\bf k}-integrated spectral function) and hybridization of the correlated Ni-$d$ orbitals with bath electrons for $Fmmm$ symmetry at -3\% strain.

\begin{figure}[!htb]
\includegraphics[width=1\linewidth]{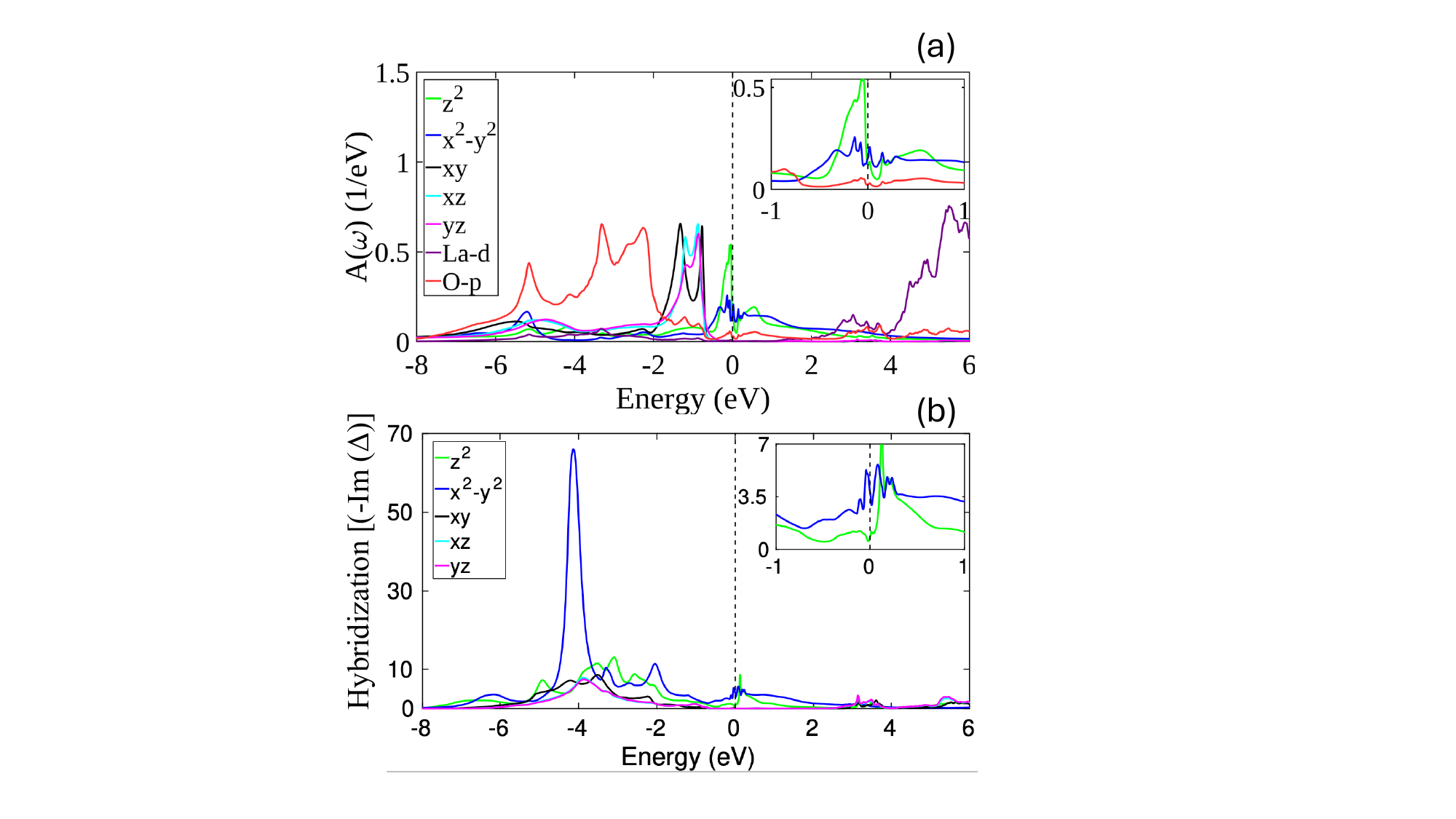}
\caption{\textbf{Results  for -3\% strains with {\it Fmmm} symmetry}. Panels (a) and (b) respectively shows the {\bf }k-integrated spectral function (partial density of states) per atom
for all the Ni-$d$, O-$p$ and La-$d$ orbitals, and hybridization of all Ni-$d$ orbitals with bath electrons. The insets are zoomed views of the two $e_g$ orbitals. Calculations were done at 300 K, with U=7 eV, J = 1 eV.}
 
 \end{figure}